\documentclass[letter,scriptaddress,superscriptaddress,twocolumn,prl,showkeys]{revtex4-2}

\usepackage{graphicx}
\usepackage{epstopdf}
\usepackage{amsmath}
\usepackage{bm}
\usepackage{epsf,epsfig,graphics}
\usepackage{float}
\usepackage{makeidx}
\usepackage{latexsym}
\usepackage{amssymb}
\usepackage{subfigure}
\usepackage{amsfonts}
\usepackage{mathrsfs}
\usepackage{dcolumn}
\usepackage{color}
\usepackage{natbib}

\begin{document}
	\title{Information Synergy in the Anticipatory Dynamics of a Retina}
	
	\author{Qi-Rong Lin}
	\affiliation{Institute of Physics, Academia Sinica, Taipei}
	\affiliation{Department of Physics, National Taiwan University, Taipei, Taiwan}
 	\author{Po-Yu Chou}
 	\author{ C. K. Chan}
	\affiliation{Institute of Physics, Academia Sinica, Taipei}
	\affiliation{Department of Physics, National Central University, Taoyuan, Taiwan}
	
	\date{\today}
	\begin{abstract}
		
       Visual perceptions often come with illusions whose physical origin are not well understood yet. The encoding of stochastic light intensity $x(t)$ into spikes with firing rate $r(t)$ at time $t$ is investigated in an experiment with retinas from bullfrogs to understand the mechanism of anticipation. Partial information decomposition of the mutual information between $r$ and the joint state $\{x,\dot x\}$ is found to be consistent with the encoding form: $r(t) \sim (1-\lambda)x(t) + \lambda \dot x(t) \tau_0$ with $\lambda$ being a system dependent parameter and $\tau_0$ a constant. This form of $r(t)$ indicates that a retina is capable of anticipation based on the synergistic information generation between $x$ and $\dot x$. Our results suggest that illusions such as the anticipation studied here during retinal perception can originate from the recombination of information extracted in the retinal network.
		
    \end{abstract}
	
\maketitle
	

%

Our visual perceptions are prone to errors which are also known as optical illusions\cite{seckel2004masters}. However, these "error" or "misinformation" are not always noises and some of them are deliberately created for useful purposes by our perception system. For example, in the phenomenon of flash-lag \cite{berry1999anticipation}, a moving object is perceived as always ahead of its actual position and thus providing more reaction time for the perceiving organism. Intuitively, information from the stimulation must be extracted, processed and/or recombined to create the illusion. This recombination or synergy of information \cite{schneidman2011synergy} is the central issue of neuroscience as it is related to how external information is being represented in our brain as perception. Unfortunately, very little is known about the physical mechanism of illusions.

Anticipation \cite{stepp2010strong}, a temporal illusion which allows animals to perceive future events, can occur as early as in the retina \cite{berry1999anticipation}. Recently, it is reported \cite{Chou2021} that the anticipatory dynamics of a retina can be understood by the physical mechanism of negative group delay (NGD) \cite{voss2016signal}; a physical concept from signal propagation in nonlinear media. In this model, the retina will generate information on its own to reflect how it perceives the future of incoming events through a delayed feedback of its past perception (experience). Arguably, the retinal circuit might be the simplest ideal network to study how information is being generated through physical mechanisms during perception.

Here we report the results of our study of the information creation process in a retina by both model simulation and experiments with frogs' retinas during the encoding of stochastic light stimulation ($x(t)$) into spikes with a time dependent firing rate $r(t)$. By assuming that $r$ contains the information created by the combination of  information from $x$ and its time derivative $\dot x$, the synergistic contribution to $r$ can be quantified by the partial information decomposition (PID) \cite{williams2010nonnegative,barrett2015exploration} of the mutual information between $r$ and the joint state $\{x,\dot x\}$. We find that a linear combination (LC) model of the form: $r(t) \sim (1-\lambda)x(t) + \lambda \dot x(t)\tau_0$ can reproduce PID results from experimental observations; with $\lambda$ being a system dependent parameter. Further comparison of the NGD model with the LC model demonstrates that the LC model also possesses NGD capabilities. Our findings suggest that there are two basic mechanisms in retinal encoding; namely relevant information extraction and their recombination.

Our experiment setup and procedures were identical to Ref. \cite{Chou2021}. Details of the experiments can be found in Refs. \cite{Chen2017,Chou2021}. Briefly, a small patch of retina from bullfrog was cut and fixed on a 60-channel multi-electrode array (MEA) from Qwane Bioscience and maintained (up to 10 hours) by the perfusion of oxygenated Ringer's solution \cite{Ishikane2005} (1 ml/min). To generate predictable signals, we used a low-pass filtered Ornstein–Uhlenbeck (OU) time series. First, the OU time series $\{s_i\}$ was generated with: $s_{i+1}  =  (1-\frac{\Delta t}{\tau})s_i + \xi_i \sqrt{D\Delta t}$ where the time step $\Delta t$ was 10 ms, $\xi$ a white noise with unit amplitude, $D = 4 s^{-1}$ the amplitude of the noise and $\tau$ the relaxation time of the system. Next, a correlated lowpass OU (LPOU) time series $\{x_i\}$ was generated from $\{s_i\}$ by using a low-pass filter with a cutoff frequency $f_c$. The illumination from an LED (peak of wavelength = 560 nm) is used to stimulate the whole retina with an intensity $I_i$ proportional to $x_i$. The maximum and minimum of $I_i$ used were $18$ and $2 mW/m^2 $ respectively with an average intensity of $10 mW/m^2$. 

Responses from the retina were then recorded by the MEA under stimulation with various $f_c$ ($1, 2, 3.5$ and $5$ Hz ) and $\tau=0.5$ sec. at $25^{\circ}C$. Since behaviors of every retina can be quite different in details and cannot be averaged, we are reporting the behaviors of a single retina below. But the reported behaviors are representative of more than 10 retinas from 10 different frogs. In a typical successful experiment, about 70\% of the MEA  electrodes are generating responses. The spikes obtained are spikes sorted to remove redundant detection. As found in Ref. \cite{Chou2021}, the responding electrodes can be classified into predicting (P-channel) and the non-predicting channels (NP-channel) because of different pathways involved in a retina \cite{famiglietti1977neuronal}. The information-theoretic analysis reported below were applied to both the response $r(t)$  and $r'(t$) from the P and NP-channels respectively. Since the anticipatory dynamics can only be found in the P-channels, the following discussions are applicable only to $r(t)$ unless otherwise noted. 

\begin{figure}[h!]
	\begin{center}
		\includegraphics[width=8.66cm]{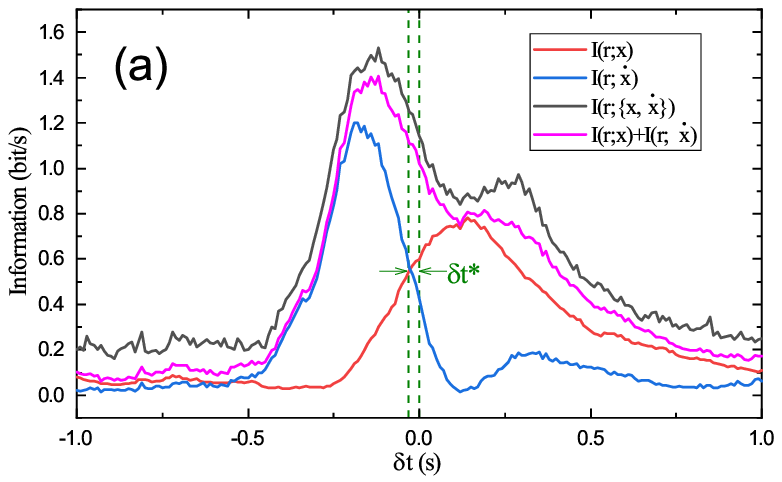}
		\includegraphics[width=8.66cm]{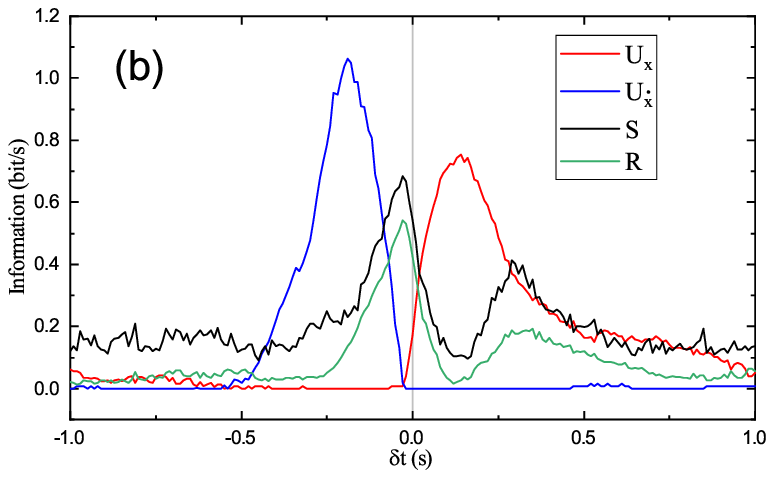}
		
	\end{center}
	\caption{Time lag dependence of a) $I(r;x)$, $I(r;\dot x)$ and $I(r;\{x,\dot x\})$ from experiments with $f_c = 1Hz$ and $\tau = 0.5 s$, b) the four components of the PID of  $I(r;\{x,\dot x\})$. Note that the $I(r;\{x, \dot x\})$ is larger than $I(r;x)$ + $I(r;\dot x)$; suggesting that there is synergy between $r$ and $\dot x$. Six states are used in the computation of mutual information. }
	\label{Expt}
\end{figure}

Similar to Refs. \cite{Chen2017,Chou2021}, three time lag mutual information (TLMI), namely $I(r;x,\delta t)$, $I(r;\dot x,\delta t)$ and $I(r;\{x,\dot x\},\delta t)$ for the mutual information of $r(t)$ between $x(t+\delta t)$, $\dot x(t+\delta t)$ and the joint state $\{x(t+\delta t), \dot x(t+\delta t)\}$ respectively can be measured to obtain the anticipatory properties of $r(t)$. To simplify notations, we will omit $\delta t$ whenever the meaning is clear from the context. 
Figure~\ref{Expt} shows these three TLMI as a function of time lag $\delta t$ in a typical experiment when $f_c$ is small enough for the retina to produce anticipatory response. One can see clearly from the figure that the peak of $I(r;x)$ is on the right of the origin; indicating that the responses (spikes) from the retina are anticipatory of $x(t)$. If $f_c$ is large enough, the peak position of $I(r;x)$ will be on the left of the origin as reported in Ref. \cite{Chou2021}. Note that there is also a peak in $I(r;\dot x)$ but its peak is located on the left of the origin. For any narrow-banded/predictable stimulation, $x$ and $\dot x$ are correlated with $\dot x(t)$ lagging behind $x(t)$. In fact, the distance between the two peaks in Figure~\ref{Expt} is the correlation time between $x$ and $\dot x$. Therefore, the retina is coding both $x$ and $\dot x$ simultaneously.

Also shown in the figure is the sum $I(r;x)$ + $I(r;\dot x)$. A remarkable feature of the figure is that this sum is smaller than the value of $I(r;\{x,\dot x\})$. That is: when $x$ and $\dot x$ are combined as a joint variable $\{x,\dot x\}$, some new information is generated  or there is synergy between $x$ and $\dot x$. In order to investigate this effect, we have followed the procedure of Beers et al. \cite{williams2010nonnegative} to perform a PID of $I(r;\{x,\dot x\})$ as:
\begin{equation}
I(r;\{x,\dot x\}) = U_x + U_{\dot x} + R + S
\end{equation}	
\noindent
where $U_x$ and $U_{\dot x}$ are the unique contributions from $x$ and $\dot x$ respectively while $R$ and $S$ are the redundant and synergistic contributions from both $x$ and $\dot x$. 

Figure~\ref{Expt}(b) shows the result of the PID. It can be seen that the peaks of $U_x$ and $U_{\dot x}$ are close to those of $I(r;x)$ and $I(r;\dot x)$ respectively as expected but there is a significant contribution to $I(r;\{x,\dot x\})$ from $S$ and $R$ in between the peaks of $U_x$ and $U_{\dot x}$. The peaks of $R$ and $S$ are located at a $\delta t^*$ such that $I(r(t);x(t+\delta t^*)) = I(r(t);\dot x(t+\delta t^*))$ in Figure~\ref{Expt}(a). In the experiments, we find that the peak positions of $R$ and $S$ are functions of $f_c$; with their peak positions shifted to the left/right with larger/smaller $f_c$. Also, their peak positions are always in between those of $U_x$ and $U_{\dot x}$ as in the case of Figure~\ref{Expt} but they can be located at either $\delta t >0 $ or $\delta t <0$. The PID of $I(r;\{x,\dot x\})$ at $\delta t = 0$ is of special interest. In the case of Figure~\ref{Expt}, it can be seen from the figure that $r(t)$ contains information from $x$ and $\dot x$ (represented by $U_x$, $U_{\dot x}$ and $R$) as well as information created by their synergy. In fact, this new information, which was not present in either $x$ or $\dot x$, is the anticipation of $x(t)$ encoded by $r(t)$. 


In a retinal circuit, the information of $x$ and $\dot x$ are presumably encoded into components of synaptic currents. Intuitively, the synaptic current responsible for the firing rate could be some linear combination  of $x$ and $\dot x$. A LC model for $r(t)$ is then $r(t) \sim (1-\lambda) x(t) + \lambda \dot x(t)\tau_0$ with $0 \leq \lambda \leq 1$. The constant $\tau_0 = 1 s$ is needed here to give correct physical dimension. The parameter $\lambda$ controls the relative importance of $x$ and $\dot x$ in the coding of $r(t)$ and needed to be determined by experiments. With this model, the encoding of $x(t)$ in $r(t)$ is not a simple one to one mapping from $x(t)$ to $r(t)$ (except for the case of $\lambda = 0$) but rather a linear  combination of $x$ and $\dot x$ after the retina has extracted $\dot x$ from $x$. To test the validity of the LC model, a numerical simulation has been performed with an encoding unit which is consisted of a threshold detection and a spike generation mechanism to give at the time step $i$:  $r_i = \mathcal{P}*\mathcal{A}((1-\lambda)x_i + \lambda \dot x_i)$ where $\dot x_i = \frac{x_i-x_{i-1}}{\Delta t}$, $\mathcal{A}(x)=max(\theta, x-\theta)$ for some threshold $\theta$ and $\mathcal{P}$ is a Poisson process with mean firing rate set to $2Hz$. 
Here we have assumed that $x$ and $\dot x$ have been extracted in the earlier layer of the retina and they serve as input to the final encoding for output. 

\begin{figure}[h!]
	\begin{center}
		\includegraphics[width=8.66cm]{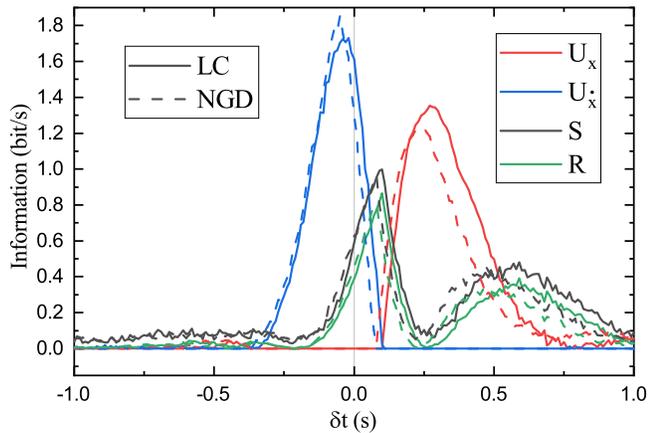}
	\end{center}
	\caption{PID of $I(r;\{x,\dot x\})$ from the simulation of the LC and NGD model to match with data from experiments shown in Figure~\ref{Expt}. Solid lines: from the LC model with $\lambda = 0.45$ and $\theta = 0$. Dotted lines: from NGD model with $\alpha = 20s^{-1}$, $\beta = {1-\lambda \over \lambda} \approx 1.22 s^{-1}$ and $kg =920s^{-2}$. $\mathcal{P}$ and $\mathcal{A}$ were applied on the outputs of both models to obtain $r(t)$. The corresponding TLMI of $I(r;x)$, $I(r;\dot x)$ and $I(r;\{x,\dot x\})$ from these two models can be found in Figure S2 in the supplementary material.}
	\label{Model}
\end{figure}

\begin{figure}[h!]
	\begin{center}
		\includegraphics[width=8.66cm]{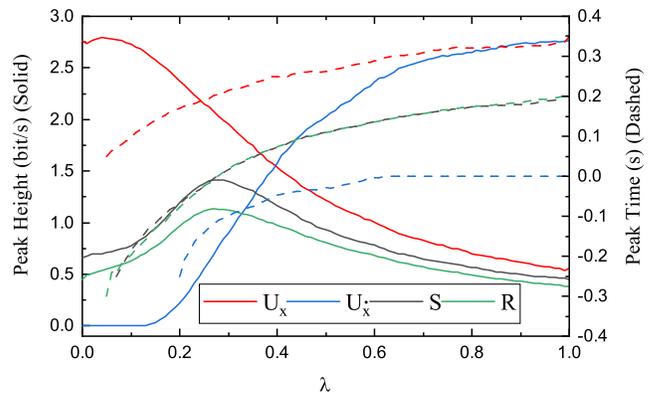}
	\end{center}
	\caption{The $\lambda$ dependence of peak heights (left y-axis; solid lines) and peak times (right y-axis; dashed lines) of $U_x$, $U_{\dot x}$, $R$ and $S$ simulated from the LC model for a stimulation with $f_c = 1Hz$ and $\tau = 0.5s$.}
	\label{Model_Prop}
\end{figure}

Figure~\ref{Model} shows the PID of $I(r,\{x,\dot x\})$ obtained from the LC model with the same stimulation used in the experiment of Figure~\ref{Expt}. In order to produce TLMI and results of PID similar to those observed in the experiment as shown in Figure~\ref{Model}, one needs to tune the value of $\lambda$. For a given stimulation, properties of the resultant PID from the LC model, such as the peak heights and peak position (in terms of time delay $\delta t$) of $U_x$, $U_{\dot x}$, $R$ and $S$ are functions of $\lambda$ as shown in Figure~\ref{Model_Prop}. Similar to the experimental observations, the peak times for $R$ and $S$ from the LC model are mostly very close and the distance (time delay) between the peak times of $U_x$ and $U_{\dot x}$ are always constant because this distance is determined only by the cross correlation between $x$ and $\dot x$. Because of these relationship between the peaks, it can be seen that all the peak positions (times) increase monotonically with $\lambda$ because $\dot x$ will the dominating the generation of $r$ as $\lambda$ increases. When $\lambda = 1$, $r$ is just $\dot x$ and therefore $U_{\dot x}$ should peak at $\delta t = 0$.  We find that the simulated TLMI and PID are not sensitive to the value of $\theta$ used in $\mathcal{A}$.

In Figure~\ref{Model_Prop}, the effects of $\lambda$ can be seen as a systematic change in the relative peak heights of $U_x$ ($h_x$) and $U_{\dot x}$ ($h_{\dot x}$); with $h_x \ll h_{\dot x}$ for $\lambda \sim 0$ while $h_x \gg h_{\dot x}$ for $\lambda \sim 1$ as $r$ will be dominated by $x$ and $\dot x$ respectively. As the shape of the PID is related to the ratio $\Pi \equiv h_x/h_{\dot x}$ and it is sensitive to $\lambda$, we use it to fix the $\lambda$ to produce PID similar to those from experiments. In fact, the $\lambda$ used to create Figure~\ref{Model} is fixed by the requirement that the $\Pi$ obtained by simulation matched that measured in the experiment. With similar procedure, we have also used the LC model to produce the PID of the experiments for other $f_c$; with results similar to those shown in Figure~\ref{Model}. An example with $f_c = 5Hz$ can be found in the supplementary material.

Although the simulation of the LC model can reproduce the shapes of different components of the PID from the experiment quite well, the positions of the PID peaks from the simulation are about 120 ms earlier than those from experiments. This time shift can be understood as the processing time of a real retina. With the Gaussian white noise spike-triggered average from the experiment, this processing time can be estimated to be about 60 ms and there is still a discrepancy of about 60 ms. In creating Figure~\ref{Model}, we could have chosen a $\lambda$ such that the peak positions are 60 ms shifted from the experimental results. But the relative shapes of the PID would be very different from those from experiments. Presumably, a more sophisticated model might be needed to account for this discrepancy.

\begin{figure}[h!]
	\begin{center}
		\includegraphics[width=8.66cm]{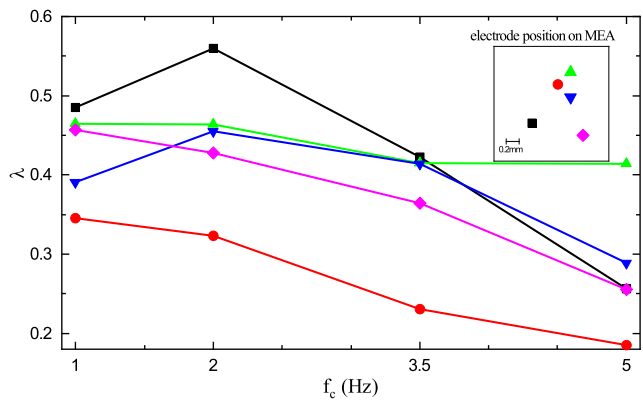}
	\end{center}
	\caption{The $f_c$ dependence of $\lambda$ in the LC model for different P-channels in the experiment. Inset: the corresponding positions of the electrodes of the P-channels.}
	\label{f_Model_Expt}
\end{figure}

Intuitively, the sensitive of a retina might be tuned to certain range of stimulation frequency. In such a case, the tuning parameter $\lambda$ should be a function of the $f_c$ used to produce the stimulation $x$ and it might also depend on the properties of the ganglion cell responsible for the firing of the P-channel. As mentioned above, there are a few P-channels for each retina. Figure~\ref{f_Model_Expt} shows the $f_c$ dependence of $\lambda$ for different P-channels in the same experiment. The positions of these P-channels in the MEA are also shown in the inset of Figure~\ref{f_Model_Expt}. There are two remarkable features in Figure~\ref{f_Model_Expt}. First, different P-channels seem to have a different $\lambda$; suggesting that different P-channels might be responsible for detection of different features from the stimulation. Second, there is a general trend of a smaller $\lambda$ for a larger $f_c$. This last observation comes from the fact that there are stronger fluctuations in $\dot x$ when $f_c$ is increased and therefore $\lambda$ needs to be smaller to accommodate the larger fluctuations in $\dot x$ at higher $f_c$. 

To relate $\lambda$ to the parameters in the NGD model in Ref. \cite{Chou2021}, we need to consider the NGD property of the LC model. If $R(\omega)$ and $X(\omega)$ are the Fourier transforms of $r(t)$ and $x(t)$ respectively, the LC model will give: $R(\omega) = H(\omega)X(\omega)$ and $H(\omega) \equiv G(\omega) e^{i\Phi(\omega)} = (1-\lambda) + i \lambda \tau_0 \omega$ with $\Phi(\omega) = tan^{-1}(\frac{\lambda \tau_0 \omega }{ 1-\lambda})$. If we require further that $\Phi$ from the LC and NGD model are the same, we have: $\lambda \approx \frac{1}{1+\beta \tau_0}$ in the parameter ranges of our experiments (Details can be found in the supplementary material). Here $\beta^{-1}$ is the time scale of the feedback variable in the NGD model. To verify this finding, the results of PID analysis of the NGD model under the same stimulation as the LC model with $\beta = {1-\lambda \over \lambda  {\tau_0}}$ are also shown in Figure~\ref{Model}. It can be seen that the PID components from the LC model and those from the NGD model are very close. The comparison of the two models shows clearly that the encoding form: $r(t) \sim (1-\lambda)x(t) + \lambda \dot x(t) \tau_0$ intrinsically possesses the NGD capability and therefore is anticipatory of $x(t)$. It might be tempting to interpret the anticipatory capability of $r(t)$ as a linear extrapolation of $x(t)$ based on $\dot x$ as: $r(t) \sim x(t) + {\lambda \over 1-\lambda} \dot x \tau_0 \approx x(t+ {\lambda \over 1-\lambda} \tau_0)$ when ${\lambda \over 1-\lambda}\tau_0$ is small. However, the values of ${\lambda \over 1-\lambda}\tau_0$ obtained are always close to $1$ which is too large for such an extrapolation picture. Therefore, the retinal circuit is producing anticipation by the recombination of the information from $x$ and $\dot x$.

As we have mentioned above, there are two types of responding channels in the MEA experiments. The results reported here are all from the P-channels. The results of TLMI and PID analysis for a typical NP-channel can be found in Figure S3 in the supplementary material. For the NP-channels, our LC model could not produce results similar to those from experiments even with negative $\lambda$. Also, we have performed PID analysis on $I(x;\{r,r'\})$ to test if there is any significant synergy between $r$ and $r'$ in the reconstruction of the input $x$. Again, we do not find that including of $r'$ can significantly improve the decoding of $x$ from $r$. The code carried by $r'$ from the NP-channels is still a mystery.

The notion that there is synergy in a neural code is not new \cite{schneidman2011synergy,latham2005synergy}. However, the interest was in how spikes from different cells can be grouped together to provide more information on the stimulation. In our case, we are interested in how different information in the simulation are recombined to create information not present in the original stimulation. The mutual information between $r$ and the joint state $\{x, \dot x\}$ indicates that it is the synergy between the $x$ and $\dot x$ that gives rise to the anticipation of the retina. That means the retinal circuit can somehow extract information of $\dot x$ from $x$ and then recombine them to form $r$. Although the phenomenological LC model cannot describe the experimental results perfectly, it captures essential features of the PID of $I(r,\{x,\dot x\})$ and it shows how the resultant recombination can be understood. Since new information is being created during this synergistic process, it should not come as a surprise that visual perceptions are prone to optical illusions. In fact, one can also consider anticipation as a form of illusion; albeit beneficial to our survival.

This work has been supported by the MOST of ROC under the grant number 108-2112-M-001-029-MY3.

\bibliographystyle{prsty}

\end{document}


\renewcommand{\thefigure}{S\arabic{figure}}

    \bibliographystyle{prsty}
    \title{Supplementary Material}
    \author{Qi-Rong Lin et al.}
    \date{\today}
{
\let\clearpage\relax
\let\newpage\relax
\maketitle
}

\section{Group Delay in the two models}
\begin{figure}[h!]
	\begin{center}
		\includegraphics[width=10cm]{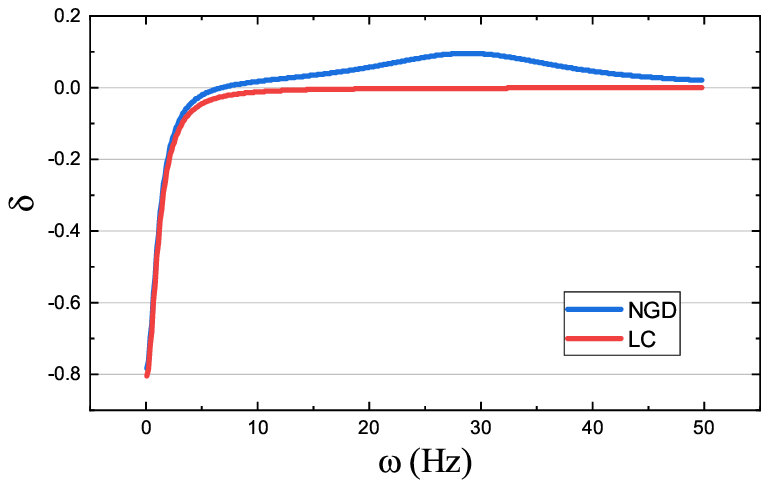}
	\end{center}
	\caption{Group delay from the two models. Parameters for the NGD model are: $\alpha = 20s^{-1}$, $\beta = 1.22s^{-1}$ and $kg =920s^{-2}$ while $\lambda = 0.45$ in the LC model. These parameters meet the two conditions so that two models can be related by $\lambda =  (1+\beta)^{-1}$ at low $\omega$.}
	\label{HeuristicVSNGD_FreqDomain_Props}
\end{figure}

The linear combination (LC) model in the main text is given by:
\begin{eqnarray}
r(t) = (1-\lambda) x(t) + \lambda \dot x(t)\tau_0
\end{eqnarray}
The NGD model in Ref. \cite{Chou2021} is given by the following equations:
\begin{eqnarray}
\begin{aligned}
\dot y(t) &= -\alpha y(t) + k[x(t)- z(t)]\\
\dot z(t) &= -\beta z(t) + g y(t)
\end{aligned}
\end{eqnarray}
\noindent
where $\alpha$ and $k$ are the relaxation time and gain of the system while $\beta$ and $g$ are defined similarly for the feedback variable $z(t)$. For the retinal model, $y$ is the cone cell while $z$ is the horizontal cell. If $X(\omega)$, $R(\omega)$ and $Y(\omega)$ are the Fourier transforms of $x(t)$, $r(t)$ and $y(t)$ respectively, the responses in the two models are:  $R(\omega)/X(\omega) = G_r(\omega)e^{i\Phi_r(\omega)}$ and $Y(\omega)/X(\omega) = G_y(\omega)e^{i\Phi_y(\omega)}$ where $\Phi_r$ and $\Phi_y$ are the phases of the responses in the LC and NGD models respectively. If we require that $\Phi_r = \Phi_y$ to provide the same group delay, we will have:
\begin{eqnarray}
\lambda = {-kg+\omega^2+\beta^2 \over [-kg + (\omega^2+\beta^2)(1-\alpha\tau_0) ] -kg\beta\tau_0}
\end{eqnarray}
\noindent
where $\tau_0=1$. It can be seen that $\lambda = (1+\beta)^{-1}$ when $kg \gg \omega^2 + \beta^2$ and $kg \gg |(1-\alpha)|(\omega^2+\beta^2)$. Since we are using low-pass OU signals with $\omega < 1$, these two conditions are met with the parameters shown in Figure 2 in the main text. Figure~\ref{HeuristicVSNGD_FreqDomain_Props} shows the comparison of the group delay obtained from these two model. It can be seen that the group delays from the two models, defined as: $\delta(\omega) = - d\Phi(\omega)/d\omega$, are close ($\delta_r\approx\delta_y$) for $\omega < 1$ and they are both negative.

\newpage
\section{Time lag mutual information (TLMI) from the two models}
\begin{figure}[h!]
	\begin{center}
		\includegraphics[width=10cm]{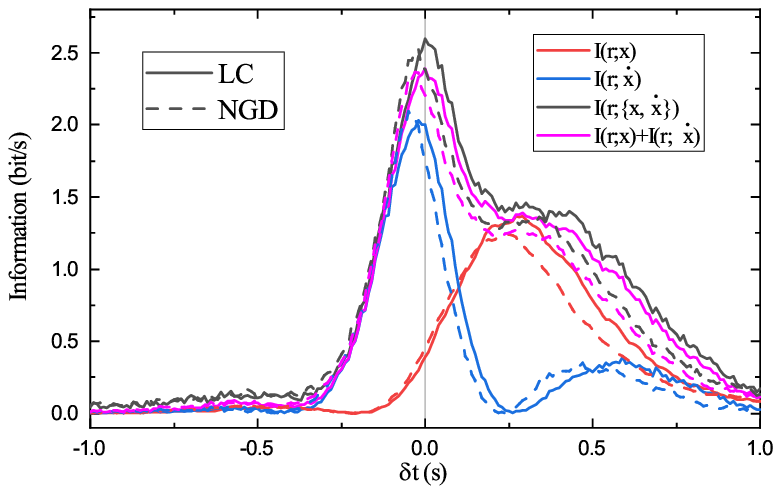}
	\end{center}
	\caption{TLMI from the LC (solid) and NGD model (dashed); with $\lambda = 0.45$, $\alpha = 20s^{-1}$, $\beta = {1-\lambda \over \lambda} \approx 1.22 s^{-1}$ and $kg =920s^{-2}$.  A lowpass filter with $f_c = 1Hz$ was used to produce the stimulation $x(t)$ from the OU signal. The same $\mathcal{P}$ and $\mathcal{A}$ used in the main text were applied on outputs of both model to generate $r(t)$. Figure 2 in the main text shows the corresponding partial information decomposition (PID). } 
	\label{HeuristicVSNGD_TLPID}
\end{figure}
\newpage
\section{TLMI and PID from experiment and LC model with $f_c=5Hz$}
\begin{figure}[h!]
	\begin{center}
		\includegraphics[width=\textwidth]{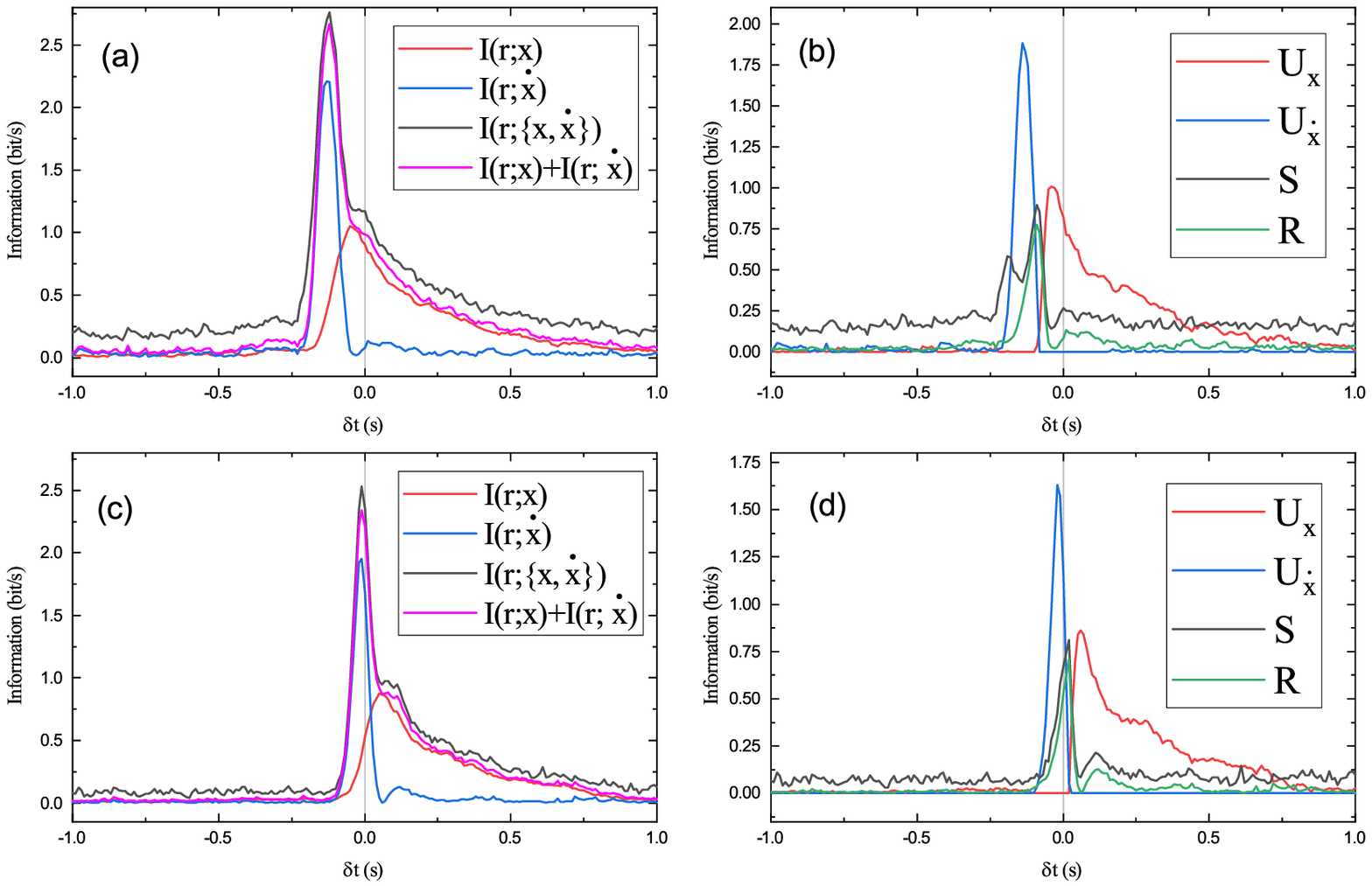}
	\end{center}
	\caption{Time lag dependence of MI and PID obtained from experiment (a,b) and the LC model (c,d) with $\lambda = 0.25$ fixed by matching the $\Pi$ from experiment with that from the simulation. The stimulation used is the LPOU with $f_c = 5Hz$.  Figure 1 in the main text shows the corresponding result from experiment with $f_c = 1Hz$. Similar to Figure 1, there is also a time shift of about 120 ms between the results from experiment and the LC model.  The time shift is considered to be the processing time of the retinal circuit. Note that this time shift is the same in the case of $f_c = 1Hz$ as shown in Figure 2 in the main text.}
	\label{5Hz}
\end{figure}

\newpage
\section{TLMI and PID from NP channel}
\begin{figure}[h!]
	\begin{center}
		\includegraphics[width=\textwidth]{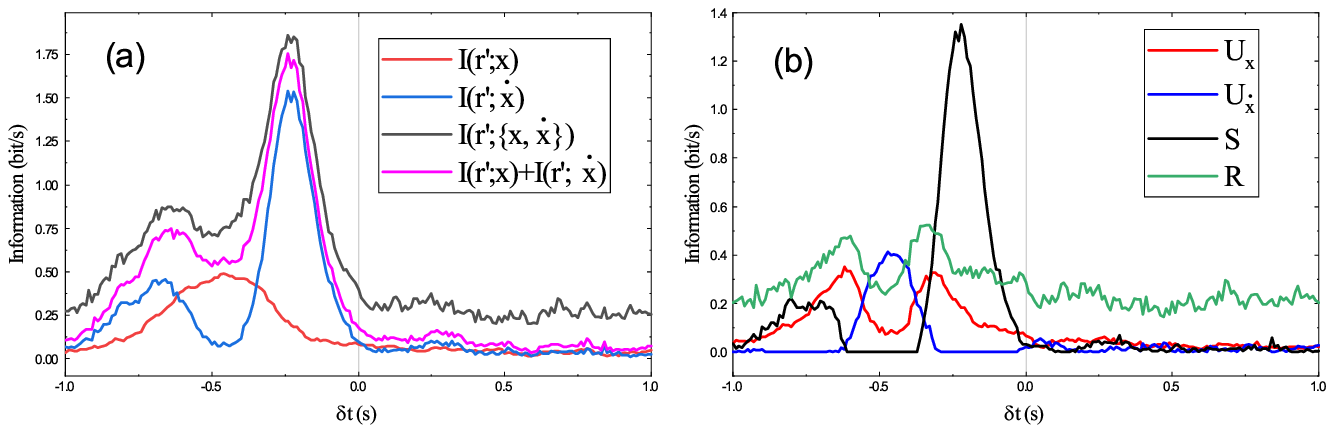}
	\end{center}
	\caption{Time lag dependence of a) $I(r';x)$, $I(r';\dot x)$ and $I(r';\{x,\dot x\})$ of a typical NP-channel and b) the corresponding four components of the PID of  $I(r';\{x,\dot x\})$. Note that data of $r'$ from an NP channel was obtained in the same experiment of Figure 1 in the main text. }
	\label{NP_TLPID}
\end{figure}

\bibliographystyle{prsty}